\documentclass[rmp,twocolumn,floatfix,letterpaper]{revtex4}

\usepackage{graphicx,amsmath,amssymb,txfonts,hyperref}

\begin{document}

\title{Model validation of simple-graph representations of metabolism}

\author{Petter Holme}
\affiliation{Department of Physics, Ume{\aa} University,
  901~87 Ume{\aa}, Sweden}
\affiliation{Computational Biology, Royal Institute of Technology,
  100~44 Stockholm, Sweden}

\begin{abstract}
  The large-scale properties of chemical reaction systems, such as the metabolism, can be studied with graph-based methods. To do this, one needs to reduce the information --- lists of chemical reactions --- available in databases. Even for the simplest type of graph representation, this reduction can be done in several ways. We investigate different simple network representations by testing how well they encode information about one biologically important network structure --- network modularity (the propensity for edges to be cluster into dense groups that are sparsely connected between each other). To reach this goal, we design a model of reaction-systems where network modularity can be controlled and measure how well the reduction to simple graphs capture the modular structure of the model reaction system. We find that the network types that best capture the modular structure of the reaction system are substrate--product networks (where substrates are linked to products of a reaction) and substance networks (with edges between all substances participating in a reaction). Furthermore, we argue that the proposed model for reaction systems with tunable clustering is a general framework for studies of how reaction-systems are affected by modularity. To this end, we investigate statistical properties of the model and find, among other things, that it recreate correlations between degree and mass of the molecules.\newline \textbf{Keywords: Chemical Networks, Complex Networks, Chemical Reaction Systems, Statistical Graph Methods}
\end{abstract}

\maketitle

\section{INTRODUCTION}

\begin{figure}
\includegraphics[width=0.85\linewidth]{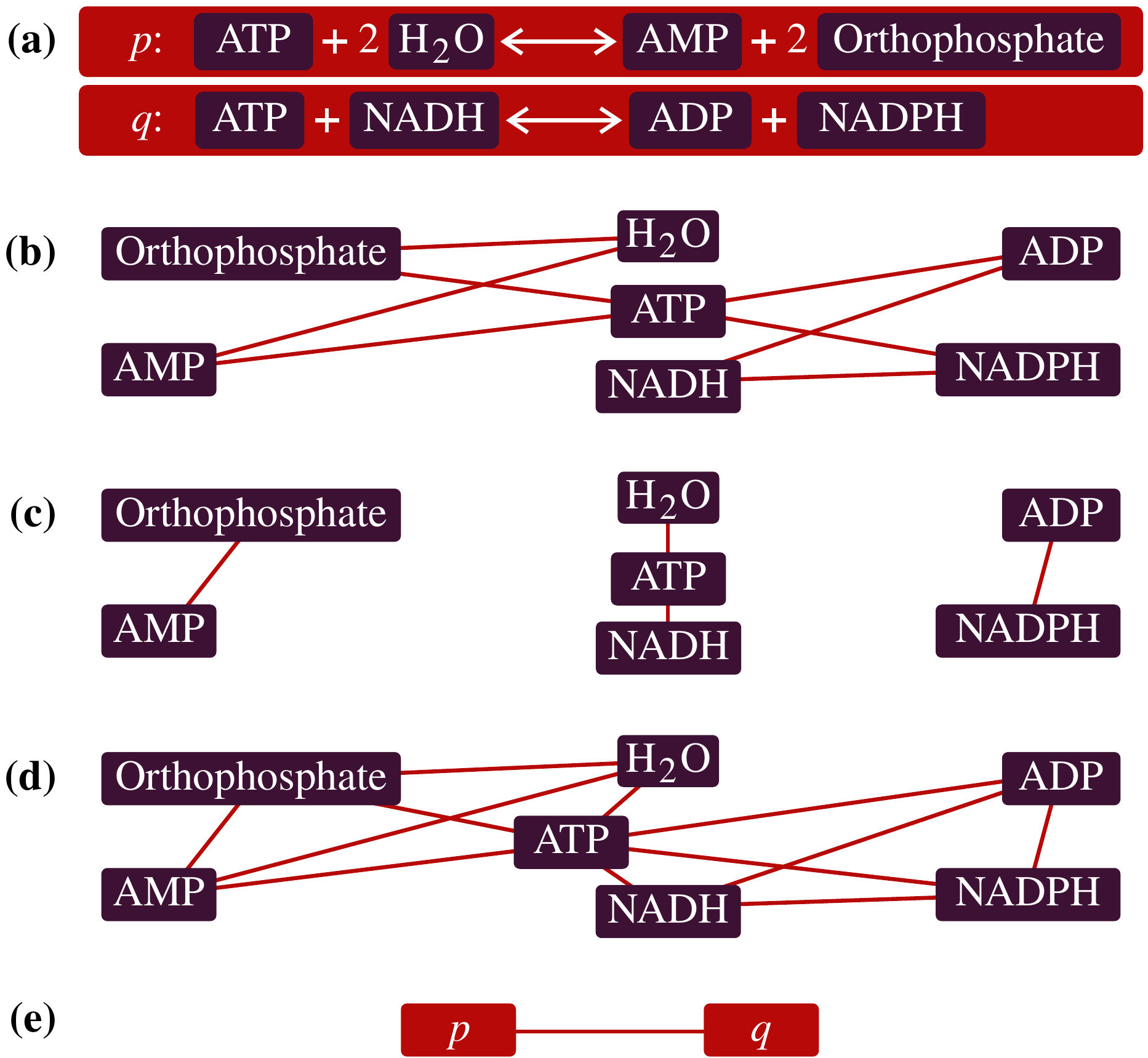}
\caption{An illustration of different network representations derived from the two hypothetical reactions shown in (a). (b) shows a substrate--product network, (c) illustrates a substrate--substrate network, (d) represents a substance network (including both the substrate--product and substrate--substrate type edges), (e) shows a reaction network where the vertices are reactions connected if they have a substance in common.
}
\label{fig:ill1}
\end{figure}

Metabolism, the set of chemical processes sustaining life in an organism, is a well-studied example of a chemical reaction system. Such systems are fundamental structures at many scales in both living organisms and the rest of the universe. From the reactions in planetary atmospheres, to the geochemical processes under the surface of our planet, and the mentioned metabolism. One can even include nuclear reactions, occurring in stars and planetary interiors, in the same framework (in this paper, however, we will use chemical terminology). Chemical reaction systems can be modelled and analyzed at different levels. At a detailed level, one can study the dynamics of the system with differential equations. Such an approach is fruitful for modelling a relatively independent subsystem, a \textit{module} (Del Vecchio 2008), such as e.g.\ the citric acid cycle. In a simple model at this level of description using mass-action kinetics, reactions are described by concentrations of the reactants, catalysts and reaction coefficients. The number of parameters (the reaction coefficients) scales like the number of reactions. For real systems, many of these parameters are unknown. (In more elaborate models, including e.g.\ temperature dependence, the situation gets even more intricate.) The complexity of modeling a large metabolic system in such a framework is staggering. The approach we take disregards all the reaction coefficients and reduces the system to a graph. Such a reduction can be done in several ways. Typically, in more elaborate graph representations (including directed edges, separating reactions, enzymes and substances, etc.)\ one can encode more information from the original system, but few general graph methods apply, so one would have to construct new ones. For simpler graph representations, more information is lost in the reduction from the reaction system to the graph, but a vast number of analysis methods are applicable.  In this work we choose the second approach and study a very simple class of graphs, conveniently termed \textit{simple graphs} --- unweighted, undirected graphs without multiple edges or self-edges. Even in such a framework, one can construct graphs from reaction systems in many ways. In Fig.~\ref{fig:ill1}, we define four types of simple graphs: \textit{substrate--product graphs} connecting the substances reacting (the substrates) with the products of the reaction, \textit{substrate--substrate graphs} linking molecules that can react with each others, or are products of the same reaction, \textit{substance graphs} connecting all substances participating in a reaction, and \textit{reaction graphs} where the vertices (nodes) are reactions and an edge represents a pair of reactions that have a common substance.  These different graph representations accentuate different aspects of the reaction system. For different questions about the system, one may need different graph representations. However, for all these representations information is lost in the conversion process. This is also true for several types of more sophisticated representations (with different types of vertices for substances and reactions, directed edges, and so on). The reason is that, with respect to the reaction dynamics, the edges are not independent. In order for mass to flow along an edge in a substrate--product graph, molecules of other vertices (than the two making up the edge) need to be present. Of course, more complex graph types can embody more information about the original system than simple graphs can. The main reason for studying simple graphs is the vast number of analysis tools that can be applied without modification.

One of the main conclusions from graph-based studies of metabolic networks is that they have a modular structure --- they can be divided into subgraphs that are more densely connected within, than between, each other (Huss \& Holme 2007, Zhao {\em et~al.}\ 2007, Zhao {\em et~al.}\ 2006, Goelzer {\em et~al.}\ 2008). Assuming that the edges of the metabolic network contribute to more or less the same degree to the dynamics of the chemical system, these network modules should be subsystems (of the dynamic reaction system) with some degree of autonomy. This is close to the general idea of biological modularity --- a biological module is commonly defined as a subsystem performing some specific, rather well defined, biological function (Del Vecchio 2008, Han {\em et~al.}\ 2004, Ihmels {\em et~al.}\ 2002, Kitano 2004). One can of course think of modules of non-biological reaction systems as well. Biological modularity is a dynamic concept that lack a precise, general definition and the link between biological and network modularity is a vast research question, beyond the full scope of this paper. Nevertheless, to be able to group and categorize modules in a biologically relevant way is important for our understanding of the biological organizations, with implications from more applied issues (like drug design) to more fundamental (like evolution). One reasonable requirement for choosing a network representation is that they should preserve the network modularity as much as possible.  In this paper, we will use this criterion, and a model of chemical reaction systems with a tunable network modularity, to investigate the above mentioned graph representations.

To outline the paper, we will start by defining the model for modular reaction systems, discuss how the network modules can be identified in simple graphs, and finally investigate how well the different graph representations can preserve the information about the modules as well as other statistical properties of the model.

\begin{figure}
\includegraphics[width=0.6\linewidth]{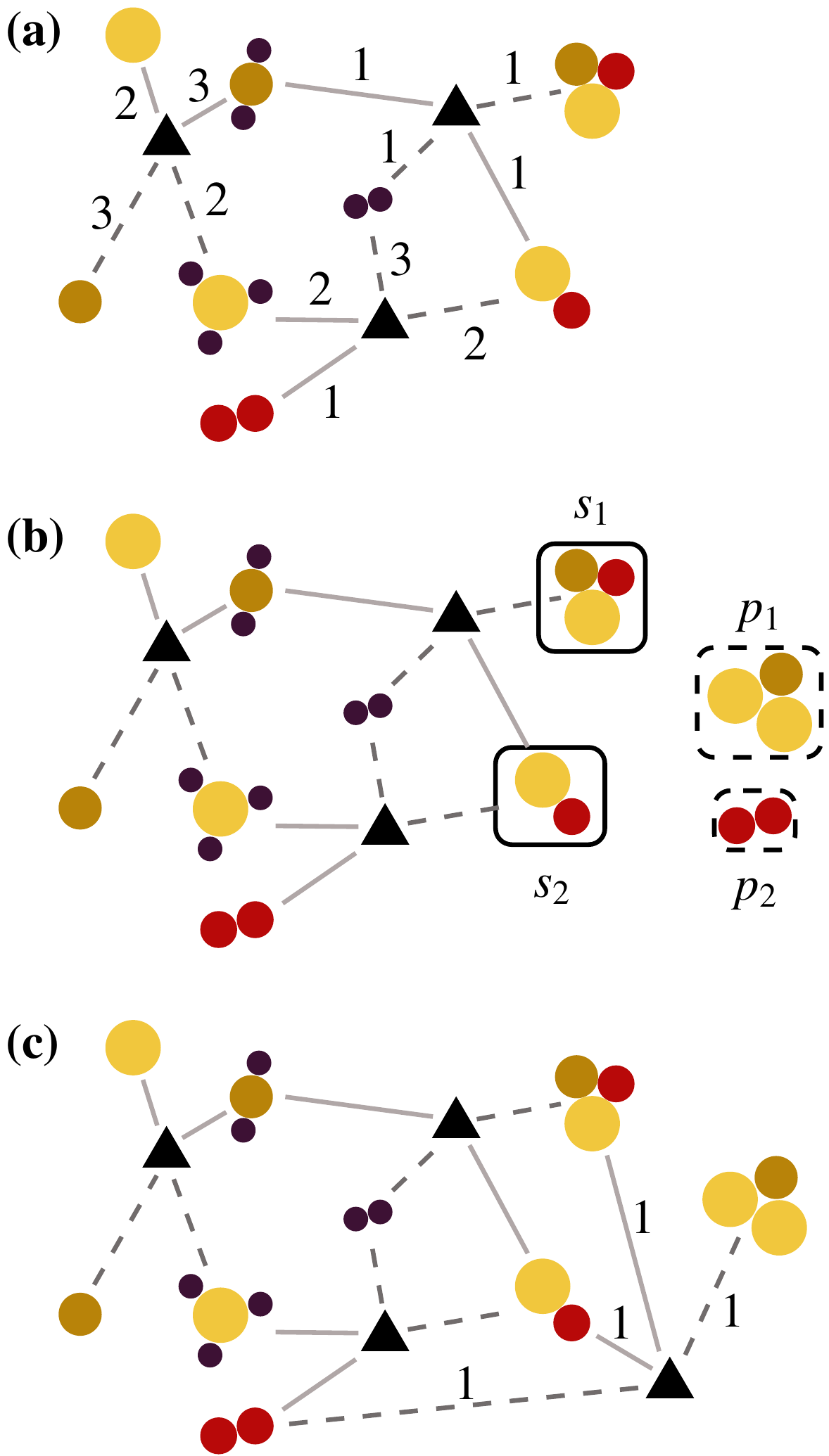}
\caption{ An illustration of the model. The circles represent atoms, grouped into molecules. Triangles represent reactions. Solid lines link substrates with reactions, while dashed lines indicate products. Numbers on the lines indicate the multiplicities $\mu_i$ of the substances in the reactions. Starting from the configuration in (a), two molecules, $s_1$ and $s_2$ in (b), are picked at random. From these, two other molecules, $p_1$ and $p_2$ and multiplicities $\mu_1$ and $\mu_2$, are generated in such a way that the mass is conserved in the reaction $\mu_1 s_1+\mu_2 s_2 \longrightarrow r_1 + r_2$. Since $p_2$ already is in the set of metabolites the resulting reaction system, from adding this reaction, looks like (c). An aspect of the model not illustrated here is that the algorithm favors products already present in the substance set. In the program we treat reactions as bidirectional, but the structure of the algorithm identical if they are directional.
}
\label{fig:ill2}
\end{figure}

\section{DEFINITIONS AND MODEL}

\subsection{The Model}

In this section, we present the model for modular chemical reaction systems mathematically. To do this we need to start by introducing a few notations. Let $X$ be a set of $N$ \textit{atoms} (in case of nuclear reaction systems ``atoms'' would mean elementary particles). A \textit{molecule} $\sigma \in \Sigma$ is a set of atoms and can be represented as a vector $\sigma = (m_1(\sigma),\cdots, m_N(\sigma))$. $m_x(\sigma)$ is the multiplicity of atom $x$ in the molecule.  A reaction is a pair of sets of molecules $(S,P)$, where $S$ is called substrates and $P$ products. $M_S(\sigma)$ denotes the multiplicity of $\sigma$ in $S$. A set of reactions is called a \textit{reaction system}.

What are the constraints on the formation of chemical reaction systems? Arguably, the most fundamental restriction is law of \textit{mass conservation}, stating that the multiplicities of all substances are the same in both $S$ and $P$ --- $M_S(\sigma)=M_P(\sigma)$ for all reactions and all $\sigma\in\Sigma$. This means that no atoms are lost, or created, during chemical reactions. A second important constraint of chemical reactions is the stability of molecules --- even if, say, an O$_4$-molecule could form by two oxygen molecules colliding, it is such a transient state that it is meaningless to count as a member of a reaction system. A third major factor shaping the reaction systems of the real world is the reaction dynamics itself. The probability of a hypothetical reaction A $+$ B $\longrightarrow$ C $+$ D  does not only depend on the presence of A, B, C and D; it also depends on the free energy change of the reaction, and the presence of catalysts (or enzymes, in a biological context) or inhibitors that can lower, or raise, the activation barrier, and thus affect the probability of a reaction. Our model will obey mass conservation, but we will keep it on an abstract and general level and not map the atoms $X$ to real atoms. Thus, the stability of molecules will not be an explicit factor in the model. In this study, reaction coefficients are not assigned to the reactions --- if needed, this could be done with an auxiliary model, maybe including correlations between the molecular mass and reaction coefficients. Furthermore, to keep the model as simple as possible we will not attempt to model other features than modularity, observed in reaction systems. One reason is that different kinds of reaction systems may have different network structure. Another reason is that some of these structures still are not completely characterized --- the degree distribution, for example, is broad, but not strictly power-law (Holme \& Huss 2008) (in fact, for the most common network representations, it is too broad to follow a power-law distribution). A third reason not to include other structure than network modularity is that it our model can easily be extended to include this.

To sketch the algorithm, let $N_A$ be the number of atoms in the system, let $N_R$ be the number of reactions, let $g$ be the number of groups (potential network modules). The algorithm starts by going through all groups adding $\gamma N_R / g$ reactions within each group. After that, the remaining $(1-\gamma) N_R$ reactions are added to tie the different modules together. If the parameter $\gamma$ is large (close to $1$), many of the reactions will take place within a group, giving the group high network modularity. In the remainder of this section, we will describe the details of these reaction addition procedures.

Consider the group $i$ ($1\leq i\leq g$). We assign the atoms $(i-1)N_A/g+1, \cdots, iN_A/g$ to this group and, furthermore, add the $N_A/g$ possible single-atom molecules to the set of $i$'s molecules $\Sigma_i$. After this, iterate the following $\gamma N_R / g$ times:
\begin{enumerate}
\item \label{step:pick} Pick two random substances $s_1$ and $s_2$ from $\Sigma_i$.
\item \label{step:assign} Assign reaction multiplicities $\mu_1$ and $\mu_2$ to these (we choose $s_1$ and $s_2$ at random, with equal probability, from the interval $[1,\cdots,\mu_{\textrm{max}}]$).
\item \label{step:create} Now, construct two potential molecules $p_1$ and $p_2$ by, for each atom $x$, assign a random fraction of the sum of multiplicities $\mu_1m_x(s_1)+\mu_2m_x(s_2)$ to $p_1$ and the rest to $p_2$ (to assure mass conservation).
\item \label{step:add} If $p_1\neq p_2$, $p_1,p_2\in\Sigma_i$, both $p_1$ and $p_2$ are different from $s_1$ and $s_2$, and the reaction $\mu_1s_1 + \mu_2s_2\longleftrightarrow p_1+p_2$ does not exist in the set of reactions of group $i$ $R_i$, then add this reaction to $R_i$. Otherwise go to step~\ref{step:pick}.
\item \label{step:if} If the algorithm has been iterated to step~\ref{step:add} $n_{\textrm{trial}}$ times without the condition that $p_1$ and $p_2$ should be in $\Sigma_i$ fulfilled, then make another iteration from step~\ref{step:pick} and add these molecules to $\Sigma_i$ and the reaction $\mu_1s_1 + \mu_2s_2\longleftrightarrow p_1+p_2$ to $R_i$.
\end{enumerate}
The purpose of the $n_{\textrm{trial}}$ iterations is to control how dense a group is. The larger $n_{\textrm{trial}}$ is, the more reactions will each molecule be involved in, i.e.\ the denser the group will be (as larger $n_{\textrm{trial}}$ means that fewer molecules have to be added to $\Sigma_i$). The reactions are assumed bidirectional, but this can trivially be changed to directed reactions. To avoid the reactions being too unbalanced with respect to mass, the $\mu_{\textrm{max}}$ parameter should be rather low (one could also think of assigning multiplicities to the product side, but to keep the model as simple as possible we do not do this. We also require there to be exactly two molecular species in both $S$ and $P$. This type of reaction is, by far the most common, making up $45\%$ of all human metabolic reactions as studied in Holme \& Huss (2008) (the second most common order is two substrates and three products, constituting $14\%$ of the reactions) or $77\%$ of the reactions in the Earth's atmosphere (Sol\'e \& Munteanu 2004). It is trivial to extend the model to sample reactions according to some distribution, empirical or not, of their order. An illustration of these steps (except the requirement that $p_1$ and $p_2$ should be in $\Sigma_i$ to break the iterations) of the algorithm can be seen in Fig.~\ref{fig:ill2}.

To add the cross-group reactions, we do as the steps above, only that we (in step~\ref{step:pick}) select substances from the entire set of present substances, not only within one group. Furthermore we skip step~\ref{step:add} (as it will be very unlikely to find $p_1$ and $p_2$ in $\Sigma$ in this case).

All in all, our model has six parameters: the number of groups $g$, the number of atoms $N_A$, the number of reactions $N_R$, the fraction of reactions within the group $\gamma$, the maximum reaction multiplicity $\mu_{\textrm{max}}$, and the number of trials to find reactions connecting already existing substances $n_{\textrm{trial}}$.

\subsection{Network modularity}

We will briefly mention how we calculate network modularity. For a more in-depth account, see Newman (2006). Consider a partition (division) of the vertex set into groups and let $e_{ij}$ denote the fraction of edges between group $i$ and $j$. The modularity of this partition is defined as
\begin{equation}\label{eq:q}
  Q=\sum_i\left[e_{ii}-\left(\sum_je_{ij}\right)^2\right],
\end{equation}
where the sum is over all groups of vertices. The term $\left(\sum_je_{ij}\right)^2$ is the expectation value of $e_{ii}$ in a random multigraph. A prototype measure for the modularity $\hat{Q}(G)$ of a graph $G$ is $Q$ maximized over all partitions, of all sizes. For metabolic networks, it is a standard practice to regard degrees as intrinsic properties of the vertices, and therefore measure network structure against a null-model $\mathcal{G}(G)$ --- random graphs with the (only) constraint that the set of degrees is the same as in $G$. We also do this, and take
\begin{equation}\label{eq:delta_q}
  \Delta(G) = \hat{Q}(G) - \langle \hat{Q}(G') \rangle_{G'\in
    \mathcal{G}(G)} ,
\end{equation}
where angular brackets denote average over $\mathcal{G}(G)$, as our measure of network modularity (Holme \& Huss 2006). We use a random rewiring of the original graph to sample $\mathcal{G}(G)$ (Maslov \& Sneppen 2002), and the heuristics proposed in Newman (2006) to calculate $\hat{Q}$. We note that it is notoriously difficult to compare modularity of networks of different sizes (Kumpula {\em et~al.}\ 2007), but, in this case, when the networks come from a series from the same network being continuously reduced, this can at most shift the peak of maximum $\Delta$ marginally (Holme \& Huss 2006).

We note that there are several other ways, apart from maximizing Eq.~\ref{eq:q}, of dividing networks into clusters. We mention the spectral methods starting from Pothen {\em et al.} (1990), and information-theory approaches, for instance, Rosvall \& Bergstrom (2007) and Ziv \textit{et al.}\ (2005), asking how the graph can best be represented as a coarse-grained graph where vertices are the clusters of the original graph in such a way that as little information as possible is lost in the process. (For even more graph clustering methods, see the references of Newman, 2006, and Rosvall \& Bergstrom, 2007.) The benefit of $Q$-maximization, as we see it, is that it is close to the notion of clusters being densely connected within and sparsely connected between each others. A cluster identified with this algorithm as the property that it cannot be divided in any way that would increase the $Q$-modularity. This give the somewhat peculiar feature that very sparse regions of the graph can be considered a cluster even if they are fragmented. For most real-world networks such regions are small and so is this effect. One common way of validating these clustering schemes is to start from a network of a number of separated cliques (fully connected subgraphs), rewire a fraction $f$ of the edges and measure the overlap between the detected clusters and the original cliques (Girvan \& Newman, 2002). Many of the proposed network clustering methods perform well in this test, which also mean they are likely to perform quite similar for the purpose of this work.

In metabolic networks, it is a common practice to delete currency metabolites --- abundant substances with a high degree that tie together modules and thereby blur the modular network structure (Huss \& Holme 2007, Holme \& Huss 2008). In our model, the new vertices introduced with the inter-module reactions do blur the modular structure, but they do not have particularly high degrees. To keep the analysis of the model simple, we do not include currency-metabolite identification.

\begin{figure*}
\includegraphics[width=\textwidth]{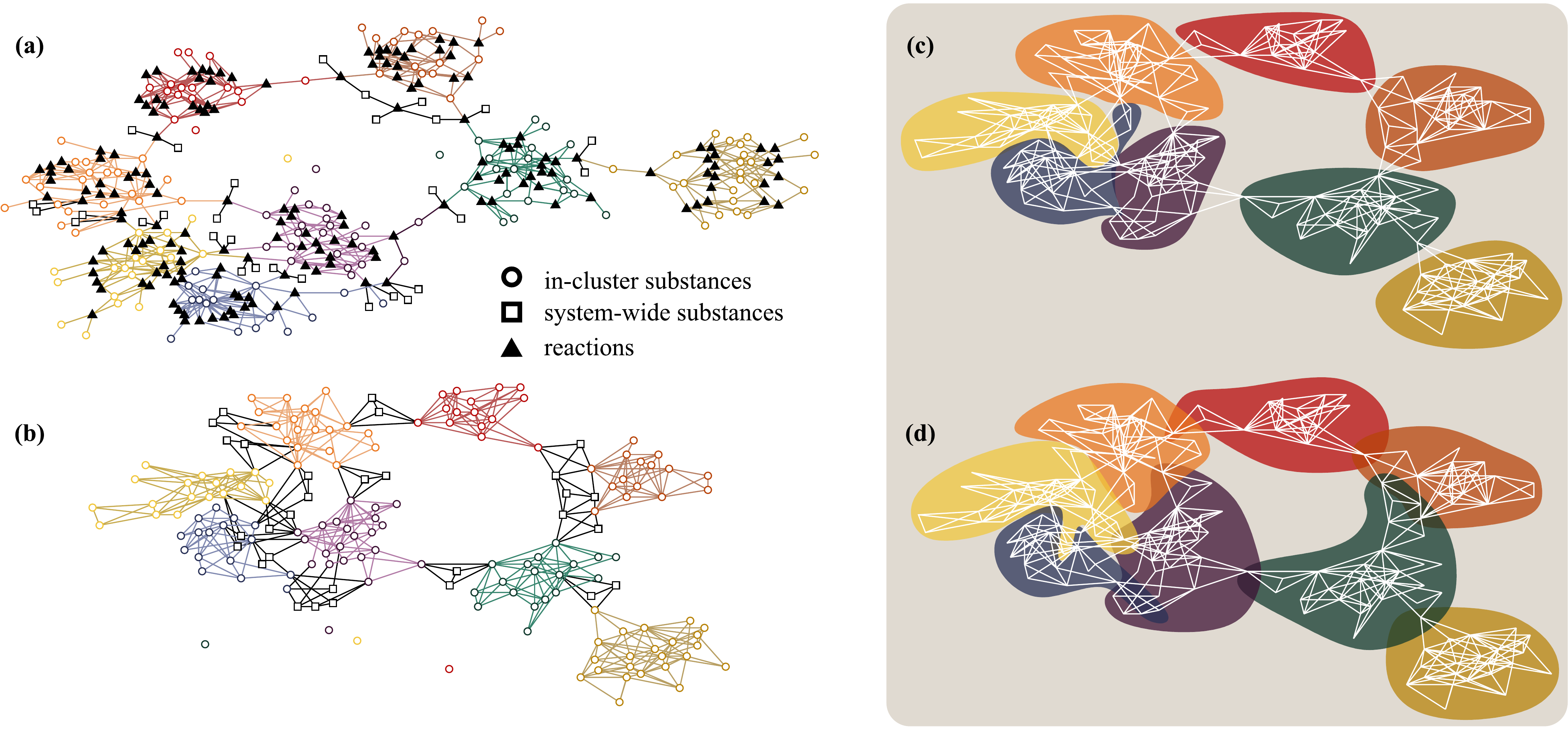}
\caption{ An example of a network realization with model parameters $g=8$, $N_A=32$, $N_R=175$, $\gamma=0.9$, $\mu_{\textrm{max}}=4$ and $n_{\textrm{trial}}=20$. (a) shows a representation of the reaction system as a bipartite network where one edge type represents reactions and the other type represents chemical substances. (b) shows the corresponding substance network. Different colors represent different groups in the construction algorithm. Vertices created in the addition of a fraction $\gamma$ of long-range reactions are symbolized by squares. (c) shows the modules detected from the substance network in (b). (d) shows the modules identified by first running the network-clustering algorithm on the reaction network, then assigning the same cluster identity to all vertices taking part in a reaction of a specific network cluster. In this case one vertex may belong to several network clusters. The underlying (white) network in (d) is, for comparison, the same as in (c).
}
\label{fig:ex}
\end{figure*}

\begin{figure}
\includegraphics[width=\linewidth]{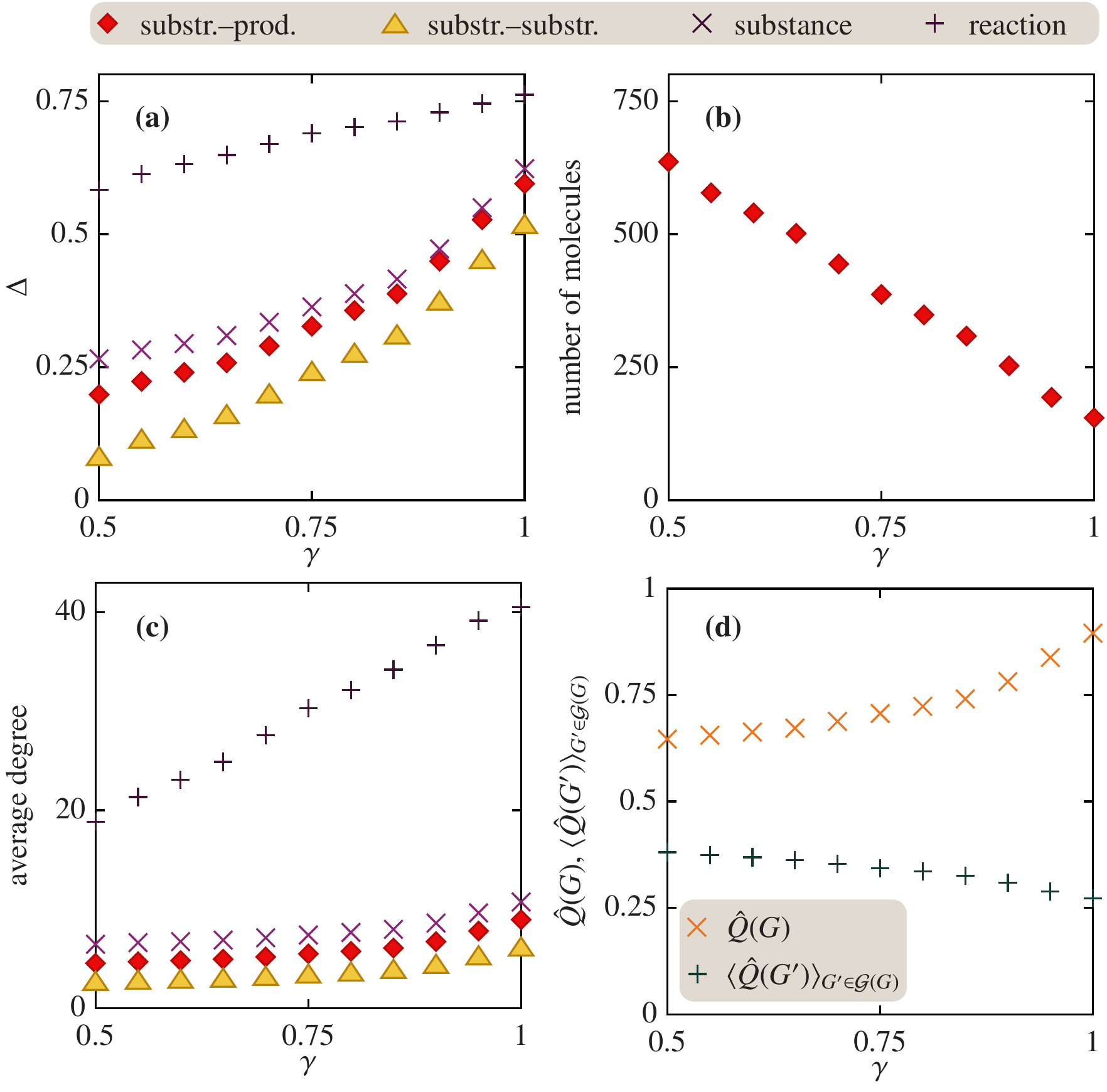}
\caption{ Statistics of the model as a function of the parameter $\gamma$. The other model parameter values are $g=10$, $N_A=50$, $N_R=500$, $\mu_{\textrm{max}}=4$ and $n_{\textrm{trial}}=100$. The points are averaged over $100$ network realizations. Standard errors are smaller than the symbols, and therefore not shown. (a) shows the relative modularity $\Delta$ of the networks as a function of $\gamma$. (b) shows the average number of molecular species for substrate--product networks (the number is, by construction, the same for substrate--substrate and substance networks), while it is exactly $N_R$ for reaction networks. (c) shows the average degree and (d) displays the two different terms in the calculation of $\Delta$ for the substance networks.
}
\label{fig:mod}
\end{figure}

\begin{figure}
\includegraphics[width=0.75\linewidth]{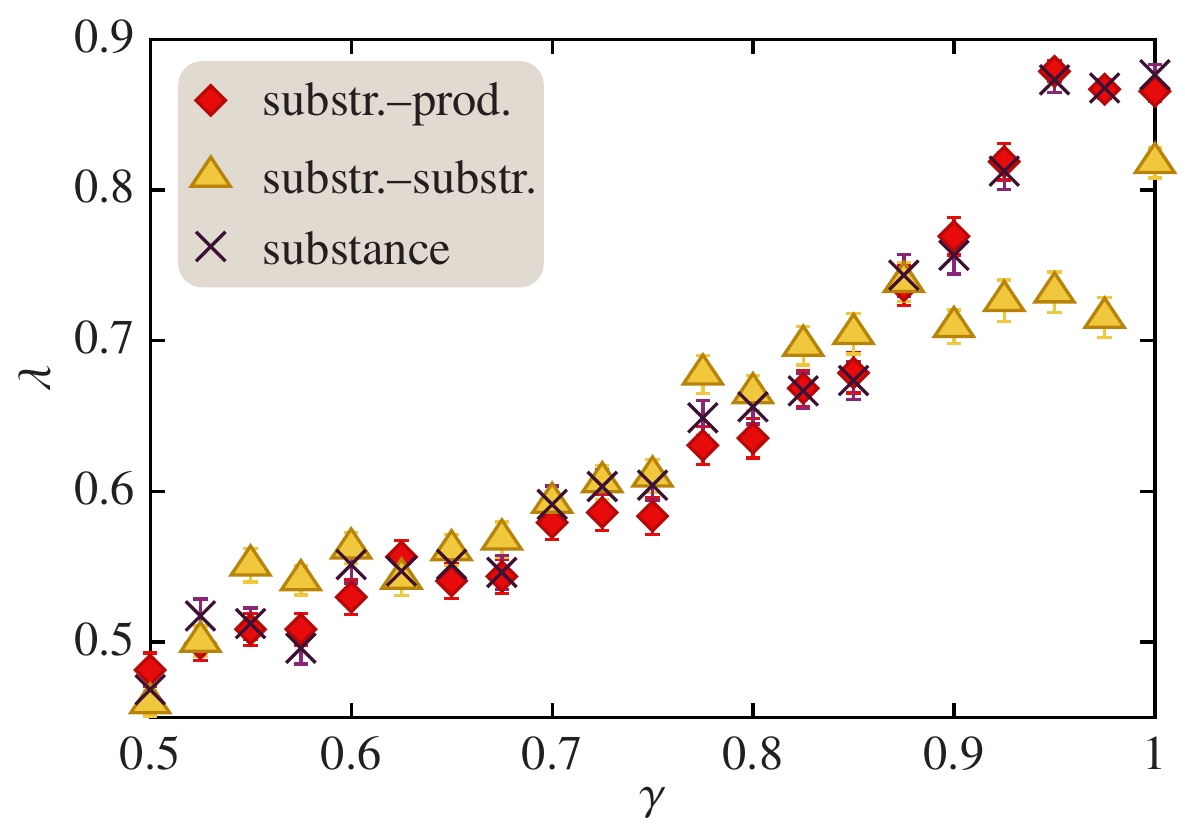}
\caption{ The predictability $\lambda$ as a function of $\gamma$. $\lambda$ is the fraction of correctly predicted group identities in matchings between the original group identities in the algorithm and the clusters identified by a network-clustering algorithm and the original groups in the network constructions.  The parameter values are the same as in Fig.~\protect\ref{fig:mod}.
}
\label{fig:qual}
\end{figure}

\begin{figure}
\includegraphics[width=0.7\linewidth]{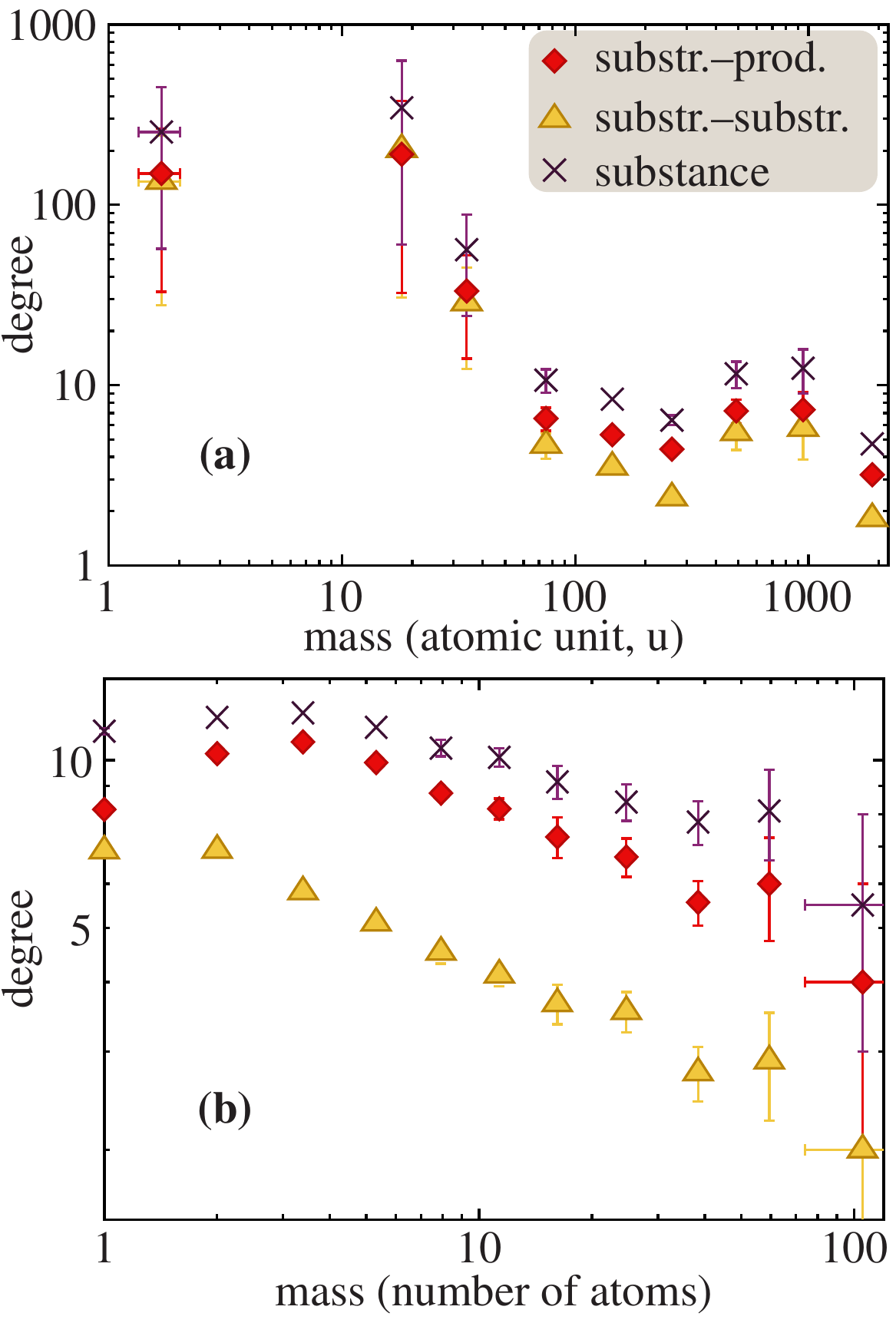}
\caption{Correlation between molecular weight and degree in networks derived from real and model reaction systems. The points are averages for logarithmic bins. (a) shows the data from the human metabolic network from the KEGG database. (b) shows the corresponding plot for our model. The mass of all atoms are set to unity in this case (so mass is just a count of the number of atoms constituting the molecule). In this figure $\gamma=0.94$ and the points are averages over $20$ networks. Other parameter values are the same as in Fig.~\protect\ref{fig:mod}.
}
\label{fig:mass}
\end{figure}

\section{NUMERICAL RESULTS}

\subsection{An example output}

In Fig.~\ref{fig:ex}(a), we display an example output of our algorithm, represented as a bipartite network where edges connect reactions (triangles) with the substances (circles or squares). The parameter $\gamma$ controlling the network modularity ($\gamma=0.90$) is rather high, giving a visually clear modular structure. The additional molecules introduced during the addition of inter-module reactions are indicated by squares. Fig.~\ref{fig:ex}(b) shows the resulting substance network. The modular structure of the bipartite representation in Fig.~\ref{fig:ex}(a) seems to be retained. From this substrate network, we run an algorithm detecting network modules (Newman 2006). This algorithm finds as many clusters as the original network, but misclassifies $\sim 2\%$ of the vertices (the additional vertices for the inter-module reactions are not counted), see Fig.~\ref{fig:ex}(c). In Fig.~\ref{fig:ex}(d) we display the output from running the modularity detection algorithm on the reaction network and assigning the identity of a module (in the set of reactions) to all vertices participating in this reaction. In metabolic databases, one metabolite can be assigned many functions; in the spirit of such multifunctionality, this kind of categorization might seem reasonable. (Note that the underlying network displayed in Fig.~\ref{fig:ex}(d) is, for comparison, the same as the substrate network of panels (b) and (c).)

\subsection{Network modularity}

$\gamma$ is the model parameter intended for tuning the network modularity of the reaction system. It is, therefore, highly desirable that the relative modularity $\Delta$ should be a monotonous function of $\gamma$ for sensible network representations and parameter values. In Fig.~\ref{fig:mod}(a) we plot $\Delta$ as a function of $\gamma$ for all types of networks discussed, and find such a monotonous relationship. Preferably, one would like the derived networks to span a large range of $\Delta$-values. All three types of networks where the vertices are substances perform comparatively well in this respect, whereas the reaction network seems a little worse (even if the actual $\Delta$-values are a larger).

When we tune $\gamma$, we keep size of the reaction system --- the number of reactions and their order (number of substrates and products) --- constant. However, the size of the derived network can be $\gamma$-dependent, which complicates the analysis of the network modularity a little. As seen in Fig.~\ref{fig:mod}(b) and (c), except the reaction network, the networks get smaller and denser as $\gamma$ is increased. The reaction network has (by construction) $N_R$ vertices constantly. Even if $\Delta=0$ represents a neutral modularity with respect to the null model, one needs to be careful when comparing $\Delta$ for networks of different sizes and average degrees.  In Fig.~\ref{fig:mod}(d) we plot the two terms, $\hat{Q}(G)$ (the maximal modularity of the network) and  $\langle \hat{Q}(G') \rangle_{G'\in  \mathcal{G}(G)}$ (the average maximal modularity of $\mathcal{G}(G)$), subtracted in the definition of $\Delta$ for the substance networks. The decrease of the second term as a function of $\gamma$ means that smaller and denser networks have lower network modularity $\hat{Q}$. If this were true for the networks derived from our reaction-system model too; then, if we choose the size of the reaction system such that the size and average degree of the derived networks are independent of $\gamma$, then the trend of growing $\hat{Q}$ would be even stronger. This indicates that $\gamma$ works as a tuning parameter for network modularity (in the substance network) whether one keeps the size of the original reaction-system or the derived networks fixed. The same conclusion can be drawn for all the other three types of networks.

\subsection{Module predictability}

As mentioned above, when $\gamma$ is large, and the network is reasonably small, one can spot the modules by inspection, like in Fig.~\ref{fig:ex}(a). By applying a network-clustering method one can then recover the group structure as in Fig.~\ref{fig:ex}(c). One desirable property of a network representation of a reaction system is that this procedure is accurate. In other words, the original, modular structure should be recovered reasonably well by the clustering algorithm in the presence of the noise created by the inter-group reactions. We measure this noise-tolerance of the network representations, or module predictability (to stress another angle of the issue), by the fraction of overlapping group identities in the best matching between the original group identities, and the identities detected by the clustering algorithm. In other words, let $x_i$ be vertex $i$'s group in the original reaction-system construction ($x_i\in [1,\cdots,g]$) and $y_i$ be vertex $i$'s identity from the graph clustering algorithm ($y_i\in [1,\cdots,h]$, $h$ is the number of detected groups). Then choose a labeling of the graph-clustering groups such that each group has a unique number in the interval $[1,\cdots,h]$, and that the number $n_{\mathrm{match}}$ of vertices $i$ with $x_i=y_i$ is as large as possible. Then we define the \textit{predictability} $\lambda = n_{\mathrm{match}} / n_g$, where $n_g$ is the number of vertices except the vertices introduced while adding inter-module reactions. We calculate $n_{\mathrm{match}}$ by a simple heuristic:
\begin{enumerate}
\item \label{step:start} Start with a random labeling of the groups.
\item \label{step:chose} Select a pair of group labels.
\item \label{step:swap} If $n_{\mathrm{match}}$ does not decrease if these labels are swapped, then swap them.
\item \label{step:rep1} If no improvement has been made during the last $n_{\mathrm{rep}}$ steps, go to step~\ref{step:chose}.
\item \label{step:rep2} Start over from step~\ref{step:start} with a new random seed unless a new highest $n_{\mathrm{match}}$ has been found in step~\ref{step:rep1} the last $N_{\mathrm{rep}}$ time steps.
\end{enumerate}
For our small system ($h$ is rarely larger than $20$), choosing $N_{\mathrm{rep}}$ and $n_{\mathrm{rep}}$ between $100$ and $10^4$ gives the same results. In Fig.~\ref{fig:qual} we display the result for the three network types with vertices being chemical substances (we do not include the reaction network, although that could, with a few additional assumptions, also be done). The matching is closest for the substrate--product and substance networks. This observation is in line with the conclusions of a study of metabolic networks (Holme \& Huss 2008) where these networks gave the most plausible set of currency metabolites and best functional overlap. Finally, we note that one major factor in the decrease of $\lambda$ as $\gamma$ is lowered from one, is that the network-clustering algorithm identifies too many groups (going up to more than twice the value of $g$). It might be the case that another network-clustering method, being more restrictive in the number of detected molecules, can give higher $\lambda$-values. In other words, the information of the original graph structure might decay slower with decreasing $\gamma$ than indicated by Fig.~\ref{fig:qual}.

\subsection{The relation between molecular mass and degree}

As a final analysis of the reaction-system model we include our explicit representation of atoms. One situation where this aspect of our model can prove useful is to explain the relation between molecular mass and degree. In Fig.~\ref{fig:mass}(a) we plot the degree as a function of molecular mass for human metabolic data from the KEGG database (same data as used in Holme \& Huss 2008). There is a decaying trend in this relationship (not clear enough to talk of a scaling law, but anyway). In Fig.~\ref{fig:mass}(b) we show the corresponding plot for our model reaction system (all network representations except the reaction networks, since the mass of a reaction is somewhat hard to define). For our model, we assume all molecules have the same weight. Neither the degree, nor the mass have such large ranges of values in the model network as in the real network. The functional form of the model curves is not matching the empirical curves; however, there is however a clear trend of decay in this data as well. Possibly, the stoichiometric constraints of the model are enough to explain this correlation in real reaction systems. More work is needed to confirm this conjecture; we leave it as a speculation in this paper.

\section{DISCUSSION AND CONCLUSIONS}

We have investigated four types of network representations of chemical reaction systems. Of these, substrate--product and substance networks are the ones that encode the information about the modularity best. This is in line with observations that these two network representations are the ones that best capture the functional organization of metabolism (Holme \& Huss 2008). To reach this conclusion, we presented a scheme to generate chemical reaction systems such that the network modularity of the derived networks can be tuned by an input parameter. The model contains molecules with atoms modeled explicitly and reactions ensured to conserve mass. The only other structure that is embedded is network modularity. One can use the model to study the response of various chemical phenomena to network modularity. It can also be a base for more elaborate modeling, including e.g.\ the highly skewed degree distributions of both metabolic networks and those derived from reaction systems in planetary atmospheres. Except some basic stoichiometric restrictions (included via the explicit modeling of the atoms), the model have very few constraints; this, however, can cause observable effects in the network topology. One such effect that we find is that our model displays a negative correlation between molecular mass and degree, a feature that also can be observed in real-world metabolic networks.

In future studies, in addition to making the model more accurate as a generative model of real reaction systems, it would be interesting to modify it to a model of the evolution of chemical reaction systems (Furusawa \& Kaneko 2006). In such an approach it would be desirable that modularity emerges from the evolutionary dynamics (Gr\"{o}nlund \& Holme 2004) rather than being imposed by the construction algorithm. Another direction to use a more dynamic interpretation of modularity (Han {\em et~al.}\ 2004, Gallos {\em et~al.}\ 2007) and find a generative model for such systems.

\acknowledgments{
  P.H. acknowledges economic support from the Swedish Foundation for
  Strategic Research and the Japan Society of the Promotion of Science and thanks Mikael Huss and Andreea Munteanu for
  data, and Mikael Huss for comments.
}

\section*{References}

{Del Vecchio}, D., Ninfa, A. \& Sontag, E., 2008 \newblock  Modular cell biology:
  {R}etroactivity and insulation.
\newblock {\em Mol. Syst. Biol.}\ {\bf 4}, 4100204.

Furusawa, C. \& Kaneko, K., 2006 \newblock  Evolutionary origin of power-laws in a
  biochemical reaction network: Embedding the distribution of abundance into
  topology.
\newblock {\em Phys. Rev. E}\ {\bf 73}, 011912.

Gallos, L.~K., Song, C., Havlin, S. \& Makse, H.~A., 2007 Scaling theory of
  transport in complex biological networks.
\newblock {\em Proc. Natl. Acad. Sci. USA}\ {\bf 104}, 7767--7751.

Girvan, M. \& Newman, M. E. J., 2002 \newblock  Community structure in social and biological networks \newblock {\em Proc. Natl. Acad. Sci. USA}\ {\bf 99}, 7821--7826 .

Goelzer, A., Brikci, F.~B., Martin-Verstraete, I., Noirot, P., Bessi\`{e}res,
  P., Aymerich, S. \& Fromion, V., 2008 \newblock  Reconstruction and analysis of the
  genetic and metabolic regulatory networks of the central metabolism of
  \textit{Bacillus subtilis}.
\newblock {\em BMC Systems Biology}\ {\bf 2}, 20.

Gr\"{o}nlund, A. \& Holme, P., 2004 \newblock   Networking the seceder model: Group
  formation in social and economic systems.
\newblock {\em Phys. Rev. E}\ {\bf 70}, 036108.

Han, J.-D.~J., Bertin, N., Hao, T., Goldberg, D.~S., Berriz, G.~F., Zhang,
  L.~V., Dupuy, D., Walhout, A. J.~M., Cusick, M.~E., Roth, F.~P. \& Vidal, M.,
  2004 \newblock  Evidence for dynamically organized modularity in the yeast
  protein-protein interaction network.
\newblock {\em Nature}\ {\bf 430}, 88--93.

Holme, P. \& Huss, M., 2008  \newblock Currency metabolites and network representations of
  metabolism.
\newblock E-print arxiv:0806.2763.

Huss, M. \& Holme, P., 2007 \newblock  Currency and commodity metabolites: Their
  identification and relation to the modularity of metabolic networks.
\newblock {\em IET Systems Biology}\ {\bf 1}, 280--285.

Ihmels, J., Friedlander, G., Bergmann, S., Sarig, O., Ziv, Y. \& Barkai, N.,
  2002 \newblock  Revealing modular organization in the yeast transcriptional network.
\newblock {\em Nature Genetics}\ {\bf 31}, 370--377.

Kitano, H., 2004 \newblock  Biological robustness.
\newblock {\em Nat. Rev. Genet}\ {\bf 5}, 826--837.

Kumpula, J., Saram\"aki, J., Kaski, K. \& Kert\'esz, J., 2007 \newblock  Limited
  resolution in complex network community detection with potts model approach.
\newblock {\em Eur. Phys. J. B}\ {\bf 56}, 41.

Maslov, S. \& Sneppen, K., 2002 \newblock  Specificity and stability in topology of
  protein networks.
\newblock {\em Science}\ {\bf 296}, 910--913.

Newman, M. E.~J., 2006 \newblock  Modularity and community structure in networks.
\newblock {\em Proc. Natl. Acad. Sci. USA}\ {\bf 103}, 8577--8582.

Rosvall, M. \& Bergstrom C. T., 2007 \newblock  An information-theoretic framework for resolving community structure in complex networks \newblock {\em Proc. Natl. Acad. Sci. USA}\ {\bf 104}, 7327--7331.

Pothen, A., Simon, H. \& Liou, K, 1990 \newblock Partitioning sparse matrices with eigenvectors of graphs. \newblock {\em SIAM J. Matrix Anal.}\ {\bf 11}, 430.

Sol\'e, R.~V. \& Munteanu, A., 2004 \newblock  The large-scale organization of chemical
  reaction networks in astrophysics.
\newblock {\em Europhys. Lett.}\ {\bf 68}, 170--176.

Zhao, J., Ding, G.-H., Tao, L., Yu, H., Yu, Z.-H., Luo, J.-H., Cao, Z.-W. \&
  Li, Y.-X., 2007 \newblock  Modular co-evolution of metabolic networks.
\newblock {\em BMC Bioinformatics}\ {\bf 8}, 311.

Zhao, J., Yu, H., Luo, J.-H., Cao, Z.-W. \& Li, Y.-X., 2006 \newblock  Hierarchical
  modularity of nested bow-ties in metabolic networks.
\newblock {\em BMC Bioinformatics}\ {\bf 7}, 386.

Ziv, E., Middendorf, M., Wiggins, C.~H., 2005 \newblock  Information-theoretic approach to network modularity. \newblock Phys. Rev. E {\bf 71}, 046117.

\end{document}